\documentclass[letterpaper]{article} % DO NOT CHANGE THIS
\usepackage{aaai25}  % DO NOT CHANGE THIS
\usepackage{times}  % DO NOT CHANGE THIS
\usepackage{helvet}  % DO NOT CHANGE THIS
\usepackage{courier}  % DO NOT CHANGE THIS
\usepackage[hyphens]{url}  % DO NOT CHANGE THIS
\usepackage{graphicx} % DO NOT CHANGE THIS
\usepackage{natbib}  % DO NOT CHANGE THIS AND DO NOT ADD ANY OPTIONS TO IT
\usepackage{caption} % DO NOT CHANGE THIS AND DO NOT ADD ANY OPTIONS TO IT
\usepackage{algorithm}
\usepackage{algorithmic}

\usepackage{xcolor}
\newcommand{\answerYes}[1]{\textcolor{blue}{#1}} 
 
\newcommand{\answerNA}[1]{\textcolor{gray}{#1}}

\usepackage{newfloat}
\usepackage{listings}
\DeclareCaptionStyle{ruled}{labelfont=normalfont,labelsep=colon,strut=off} % DO NOT CHANGE THIS
\floatstyle{ruled}
\newfloat{listing}{tb}{lst}{}
\floatname{listing}{Listing}
\usepackage{subcaption}
\usepackage[colorlinks=false,hidelinks=true]{hyperref}

\title{SocioXplorer: An Interactive Tool for Topic and Network Analysis in Social Data}
\author {
    Sandrine Chausson,
    Youssef Al Hariri,
    Walid Magdy,
    Björn Ross
}
\affiliations {
    The University of Edinburgh, School of Informatics \\
    b.ross@ed.ac.uk
}
\usepackage{bibentry}
\begin{document}

\maketitle

\begin{abstract}
SocioXplorer is a powerful interactive tool that computational social science researchers can use to understand topics and networks in social data from Twitter (X) and YouTube. It integrates, among other things, artificial intelligence, natural language processing and social network analysis. It can be used with ``live" datasets that receive regular updates.
SocioXplorer is an extension of a previous system called TwiXplorer, which was limited to the analysis of archival Twitter (X) data. SocioXplorer builds on this by adding the ability to analyse YouTube data, greater depth of analysis and batch data processing.
We release it under the Apache 2 licence\href{https://github.com/smash-edin/socioxplorer}.

%TwiXplorer . However, as the need for alternative sources of data is growing, the tool's focus on Twitter data limits its usefulness. Moreover, it is currently impractical to use it with  and its features remain somewhat isolated from each other, hence limiting the depth of analysis it enables. We therefore present SocioXplorer, an extension of TwiXplorer that supports both Twitter and YouTube datasets, enhances the tool's analytical capabilities thanks to new and improved features, and allows batch data integration. We make this tool fully accessible.
\end{abstract}


\section{Introduction}

Social media platforms are critical sources of data for researchers, and Twitter (X) in particular has garnered significant attention. As a result, many tools have been developed to facilitate the exploration of Twitter datasets. One such tool is TwiXplorer \cite{al2024twixplorer}. Accessing Twitter data, however, has become increasingly challenging \cite{twitter_api} and an over-reliance on Twitter data has long raised concerns over skewed representations of social realities \cite{Karakikes2024BiasIX, Küpfer_2024}.

YouTube, on the other hand, stands out as a rich yet under-explored data source \cite{caren2020contemporary, Bartolome2023}. With more than two billion active users per month \cite{youtube_statista}, it has for instance been used to study misinformation \cite{rochert2021networked}, opinion formation \cite{rochert2020opinion}, radicalisation \cite{ribeiro2020auditing}, or social activism \cite{caren2020contemporary}. Yet, very little research has been dedicated to developing tools for the analysis of YouTube datasets.

With this in mind, we present SocioXplorer, an extension of TwiXplorer capable of handling YouTube data as well as Twitter data.  SocioXplorer also improves on some of TwiXplorer's limitations. Namely, it introduces new features that enhance the tool's analytical capabilities and enables a more flexible integration of data into the system through batches. Together, these additions significantly extend the usefulness of the original tool, while allowing for a more insightful exploration of data. We make SocioXplorer freely accessible with detailed installation instructions and demos\footnote{\href{https://github.com/smash-edin/socioxplorer}{https://github.com/smash-edin/socioxplorer}}.

% TwiXplorer was also originally designed for the analysis of archival data, making it impractical to use when data is still being collected. Finally, while useful, TwiXplorer's visualisation features remain somewhat siloed. For instance, while the user can easily identify topics and communities in the data, these insights cannot easily be combined.

\section{System description}

TwiXplorer tool was originally designed to analyse archival Twitter data. For a given set of search criteria, the tool generates a \textbf{timeline} of the data, the distribution of \textbf{languages} and \textbf{sentiments}, a map of tweets' and users' \textbf{geolocation}, a table of the \textbf{most popular content} (i.e. posts, images and URLs) and \textbf{wordclouds}. The tool also contains two more ``advanced" features. The \textbf{Topic Discovery} allows the user to visualise the dataset in semantic space, locate specific claims and cluster tweets into topics. The \textbf{Social Network Analysis (SNA)} component, on the other hand, shows users' proximity to each other based on their retweets and clusters them into communities using this information. Finally, TwiXplorer is highly interactive: the user can easily dig into the data by clicking on visualisations, these are zoomable and filterable and tweets can be viewed in context on Twitter directly. Please see the Appendix for features screenshots.

\subsection{Extension 1: YouTube data support}

TwiXplorer offers a range of powerful features that can help researchers carry out data analysis. However, the tool's focus on Twitter data only limits its usefulness. The main way in which SocioXplorer extends on TwiXplorer is therefore by providing support for the analysis of YouTube data. This is achieved with minimal disruption to the UX logic of TwiXplorer by making YouTube comments the main unit of analysis, similarly to tweets in the context of Twitter.

Replying is the main mode of interaction on YouTube. Importantly, the platform does not support ``chained replies": i.e. one can comment on the video itself, or reply a comment. However, they cannot reply to a reply. Users often work around this constraint by adding their reply to the thread and mentioning the account they are replying to.

This has implications when adapting the SNA for YouTube data. In our implementation, we treat ``top-level" comments as replies to the channel. Replies to top-level comments, if they do not mention authors of previous replies, are replies to authors of these comments. Otherwise, they are replies to the repliers. Channels and commenters are both treated in the same way as users.

Note that, while Twitter data sometimes includes geolocation information, YouTube data does not. This means that reports generated by SocioXplorer for YouTube datasets do not include the ``geolocation" visualisation.

\subsection{Extension 2: Enhanced analytical capabilities}

While powerful, the usefulness of some of TwiXplorer's visualisations is limited by information being somewhat siloed. For instance, while the user can easily identify topics and communities in the data, these insights cannot easily be combined. SocioXplorer addresses this limitation thanks to new and improved features. 

\begin{figure}[h!]
    \centering
    \includegraphics[width=\linewidth]{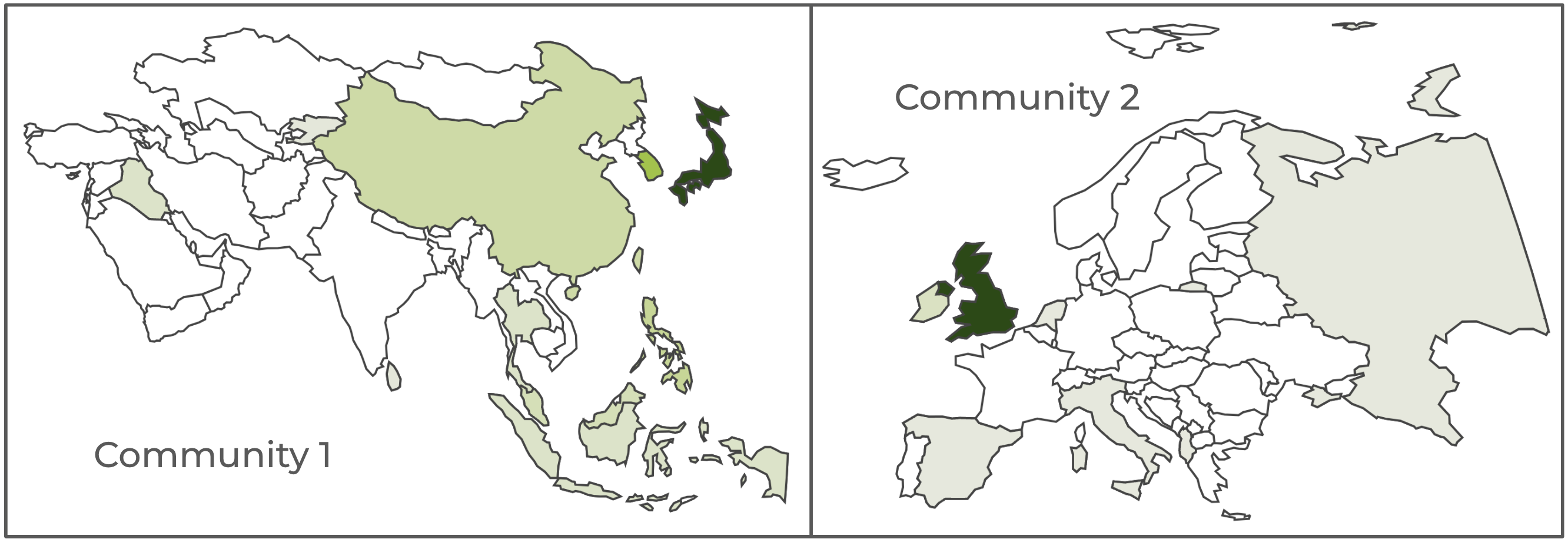}
    \caption{Geolocation component after using the ``Community" filter to focus on specific communities.}
    \label{fig:coms_per_country}
\end{figure}

The first of these is a ``community" filter added to the language, geolocation and topic discovery visualisations. By linking community information with these other dimensions, SocioXplorer makes it easier for the researcher to characterise communities. For instance, Figure \ref{fig:coms_per_country} shows the spatial distribution of two communities in a 1M tweets dataset related to the 2022 FIFA World Cup and reveals they are predominantly located in Japan and the UK respectively. 

A second addition allows the user to label communities and topics, thus enabling them to record insights directly into the tool while improving the readability of visualisations. For instance, after looking at user descriptions, individual accounts, geolocation and language distributions, we chose to label the communities from \ref{fig:coms_per_country} as ``K-pop fans" and ``UK football news" respectively.

\begin{figure}[h!]
    \centering
    \includegraphics[width=\linewidth]{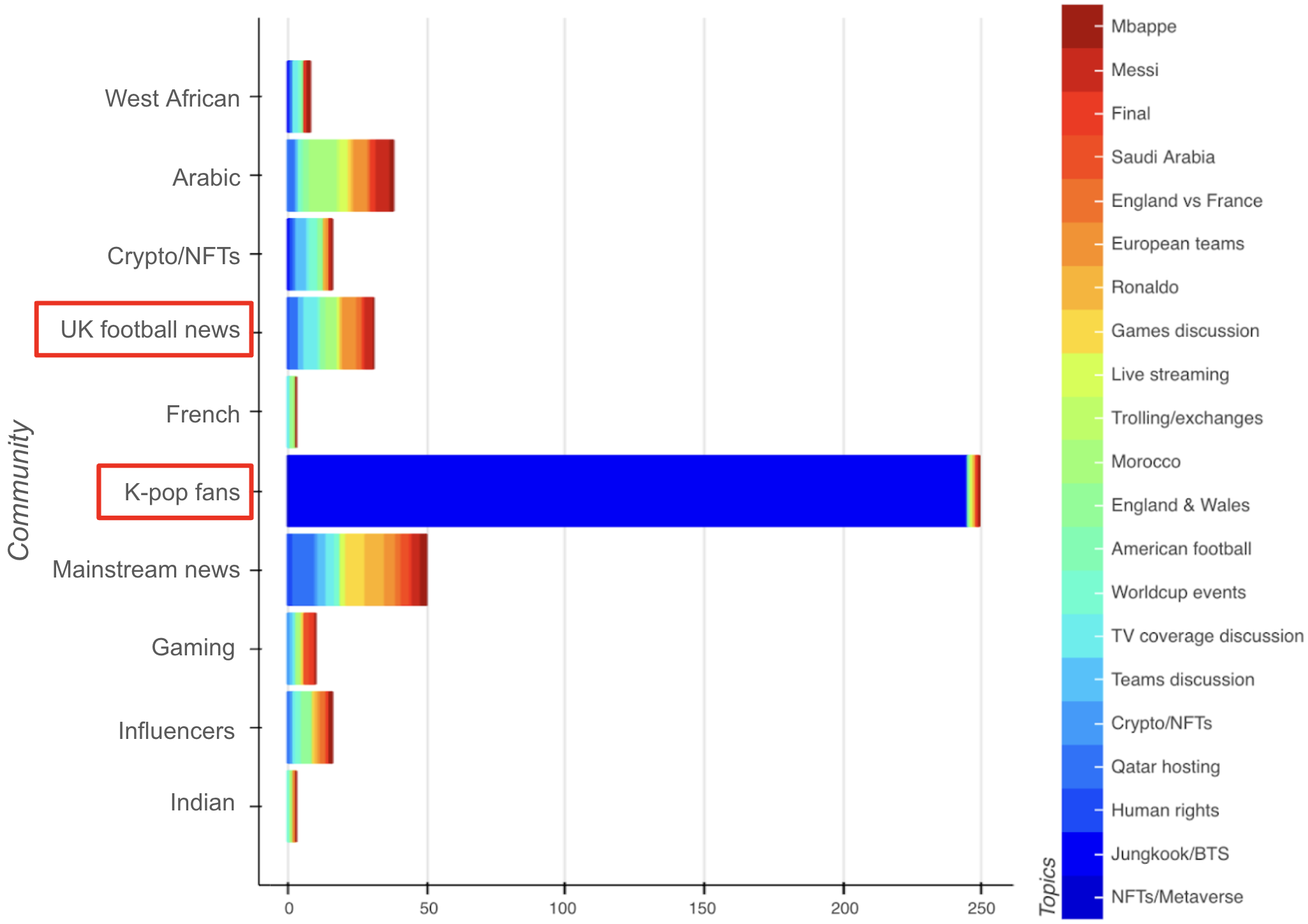}
    \caption{Topics per community in 2022 FIFA World Cup Twitter dataset.}
    \label{fig:topics_per_com}
\end{figure}

Finally, the third addition is a visualisation showing the distribution of topics per community (as either counts or proportions). This allows the researcher to investigate which communities are engaging with which topics of conversation. For instance, Figure \ref{fig:topics_per_com} reveals that the ``K-pop fans" community engaged almost exclusively with the ``Jungkook/BTS" topic unlike the ``UK football news" community, who engaged with a wider range of topics. 

\subsection{Extension 3: Batch processing}

The original TwiXplorer tool was designed for datasets that are fully collected by the time they are analysed. SocioXplorer, on the other hand, allows researchers to incorporate new batches of data into the system, thus accommodating for the analysis of incrementally collected data. This ``batch processing" extension, however, comes with challenges with respect to the SNA, which needs to remain visually consistent when incorporating new data. To achieve this, the positions of existing users is loaded from memory while that of new users is randomly initialised. In practice, this leads to new users quickly converging to their optimal location when applying the Force Atlas 2 algorithm \cite{jacomy2014forceatlas2} while limiting the displacement of existing users. 

\begin{figure}[h!]
    \centering
    \includegraphics[width=\linewidth]{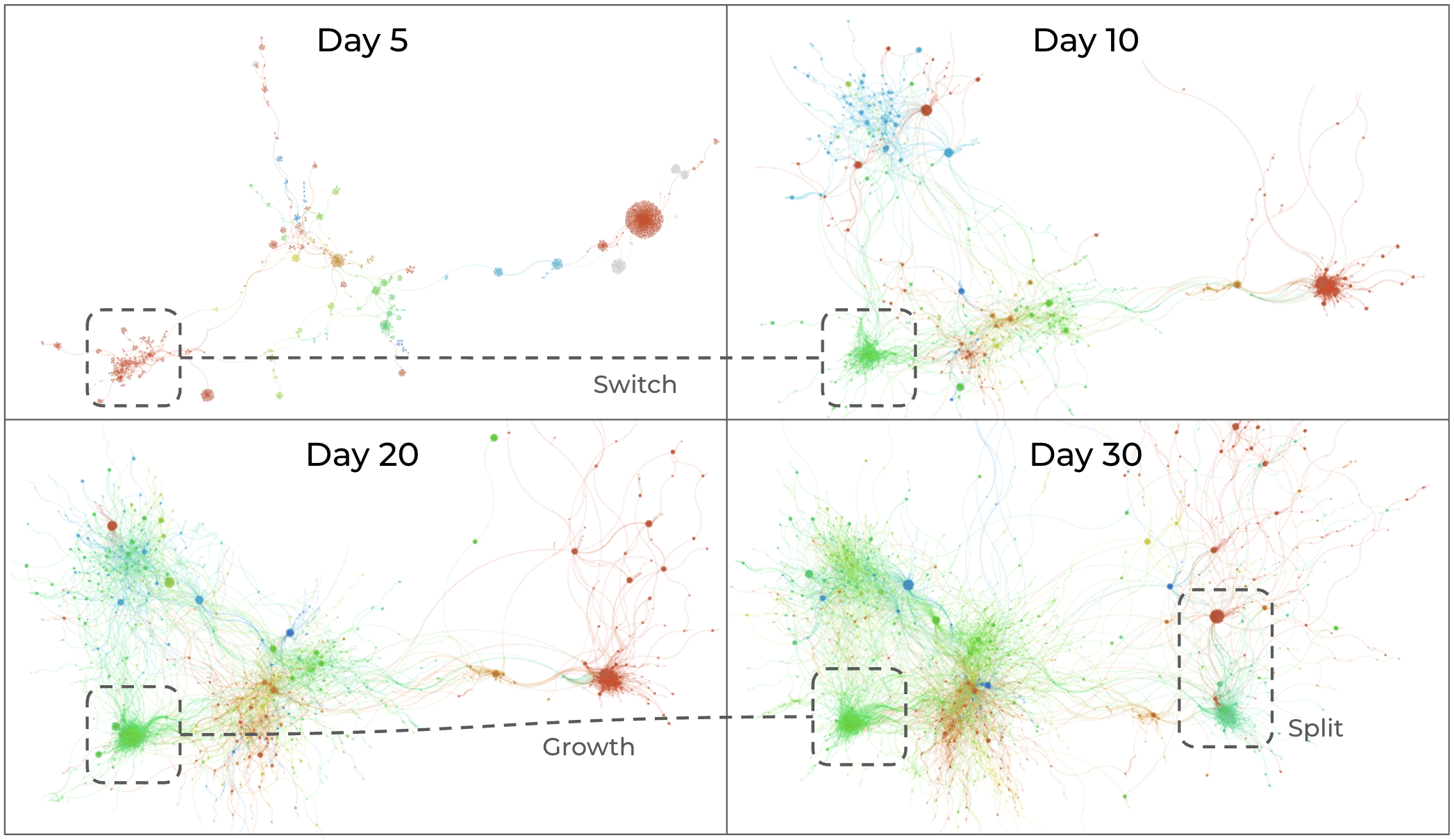}
    \caption{Snapshots of reply network from YouTube dataset as obtained through daily batches. Dots are users and colours are communities. These snapshots show users switching between communities (``Switch"), communities growing larger (``Growth") and communities splitting (``Split").}
    \label{fig:batching_coms}
\end{figure}

Moreover, to preserve community labels when new data is added, we cluster users into communities from scratch using the Louvain algorithm \cite{blondel2008fast} and map these new communities to old ones based on their overlap. If this overlap is greater than a given threshold, the new community inherits the same label as the old community. Otherwise, it is treated as a new community altogether. Figure \ref{fig:batching_coms} illustrates this batch processing for the SNA on a YouTube dataset of about 4M comments related to climate change.

\section{Conclusion}

In this paper, we presented SocioXplorer, an extension of the TwiXplorer tool that supports Twitter as well as YouTube data, introduces more powerful analytical capabilities thanks to new visualisation, filtering and labelling features, and supports the integration of data collected incrementally. We hope that SocioXplorer can facilitate researchers' work as well as encourage them to use YouTube data.

\section{Ethics Statement}

All external libraries used to develop SocioXplorer are covered by an open-source license authorising their use for research, i.e. Apache-2.0 License or BSD 3-Clause License. SocioXplorer itself is released publicly under the Apache-2.0 License, so that researchers can use it for the analysis of their own data.

SocioXplorer was developed on the authors' portable computers and a single desktop computer hosting the datasets. It was tested using a dataset of up to ~4M YouTube comments and ~45M tweets. A GPU was used for pre-processing steps related to Sentiment Analysis and Topic Discovery.

Given the potential inclusion of personal information in Twitter and YouTube, we cannot guarantee that users won't use SocioXplorer to identify individual users or process personal information. None of the features of SocioXplorer were specifically designed to enable this though, but rather to generate aggregate insights about the data. Moreover, there is a pressing need to better understand narrative spread on social media (e.g. misinformation, hate speech, astroturfing) and we believe the benefits of our tool for research outweigh any risk of misuse.

Finally, our system will be a useful addition to the toolbox of anyone trying to understand multifaceted social media communication. However, we emphasise that is not designed to replace human analysts or qualitative research, but rather to enhance and facilitate it.

\section{Acknowledgments}
This work was supported by the Turing's Defence and Security programme through a partnership with the UK government and Dstl in accordance with the framework agreement between GCHQ \& The Alan Turing Institute, the UKRI Centre for Doctoral Training in Natural Language Processing (grant EP/S022481/1) and the School of Informatics at the University of Edinburgh.

\bibliography{aaai25}

\newpage

\noindent\begin{minipage}{\textwidth}

    \section{Appendix}
    
    \begin{figure}[H]
        \centering
        \begin{subfigure}[b]{0.45\textwidth}
            \centering
            \includegraphics[width=\textwidth]{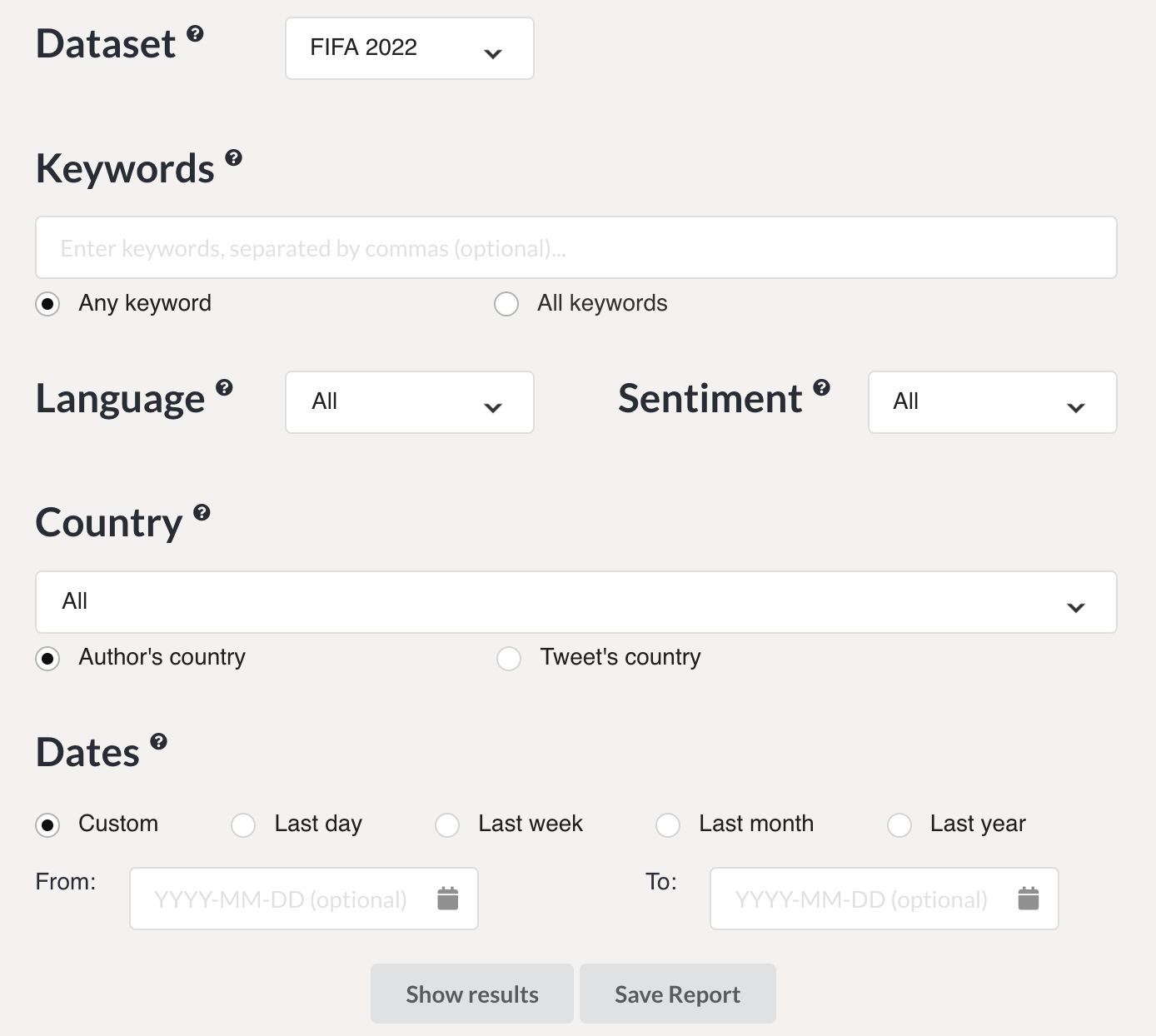}
            \caption{Input fields}
            \label{fig:input}
        \end{subfigure}
        \hfill
        \begin{subfigure}[b]{0.45\textwidth}
            \centering
            \includegraphics[width=\textwidth]{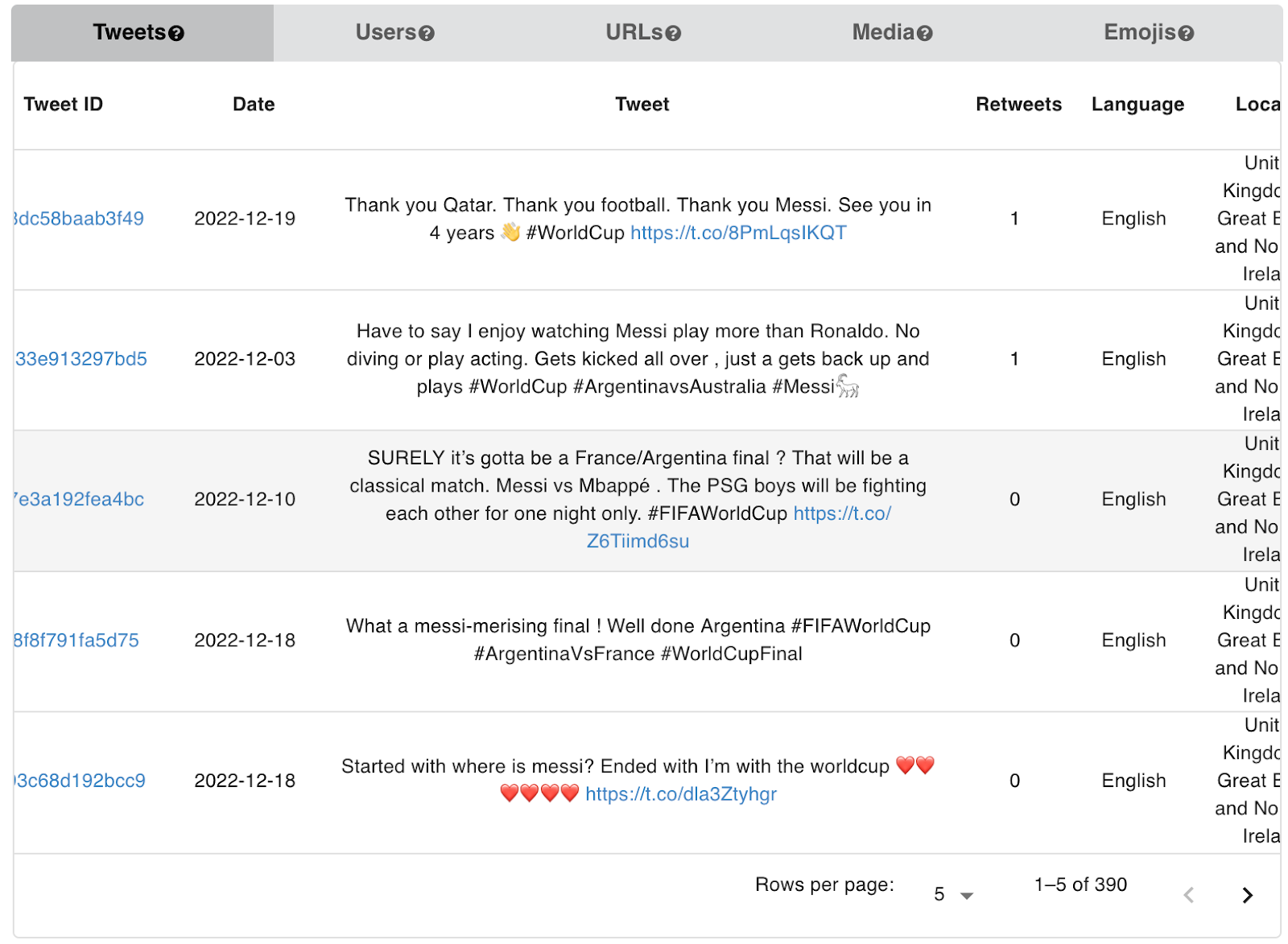}
            \caption{Top content}
            \label{fig:top_content}
        \end{subfigure}
        \label{fig:sidebyside}
    
        \vspace{1em}
        
        \begin{subfigure}[b]{0.45\textwidth}
            \centering
            \includegraphics[width=\textwidth]{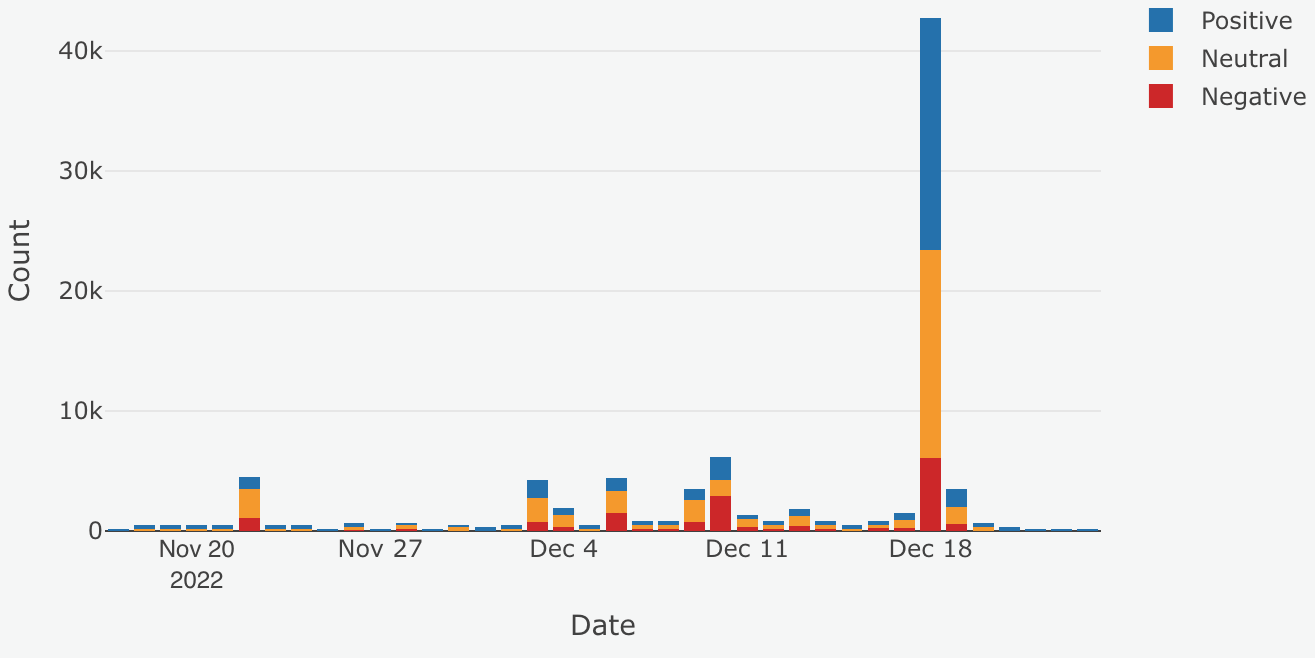}
            \caption{Sentiment timeline}
            \label{fig:sentiment}
        \end{subfigure}
        \hfill
        \begin{subfigure}[b]{0.45\textwidth}
            \centering
            \includegraphics[width=\textwidth]{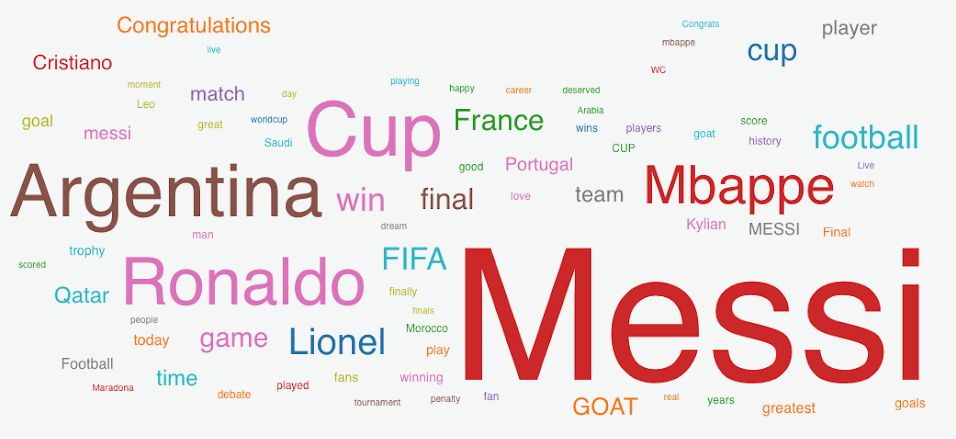}
            \caption{Wordcloud}
            \label{fig:wordcloud}
        \end{subfigure}
        \label{fig:sidebyside}
    
        \vspace{1em}
    
        \begin{subfigure}[b]{0.45\textwidth}
            \centering
            \includegraphics[width=\textwidth]{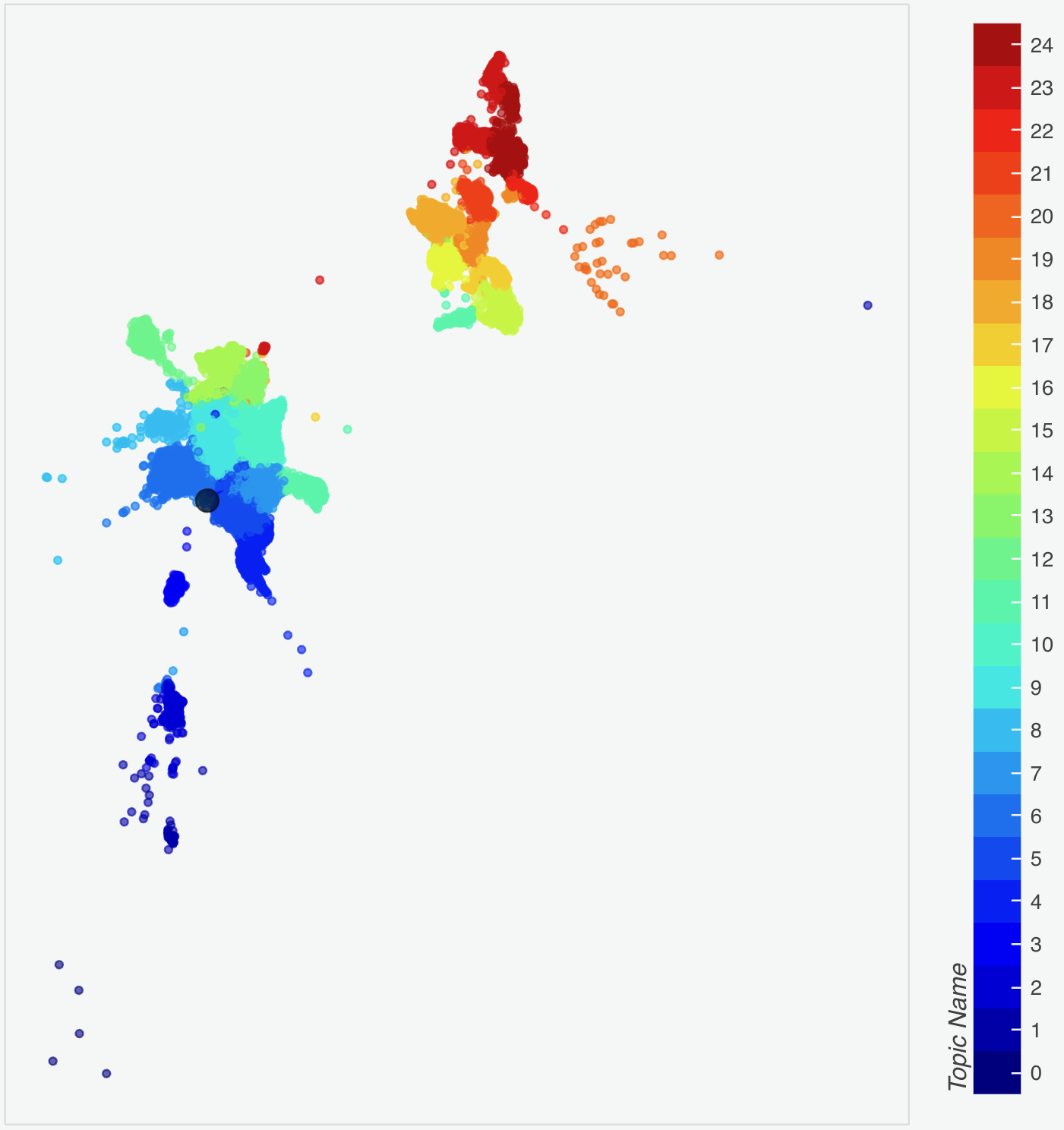}
            \caption{Topic Discovery with claim detection (large black dot).}
            \label{fig:topic_disc}
        \end{subfigure}
        \hfill
        \begin{subfigure}[b]{0.45\textwidth}
            \centering
            \includegraphics[width=\textwidth]{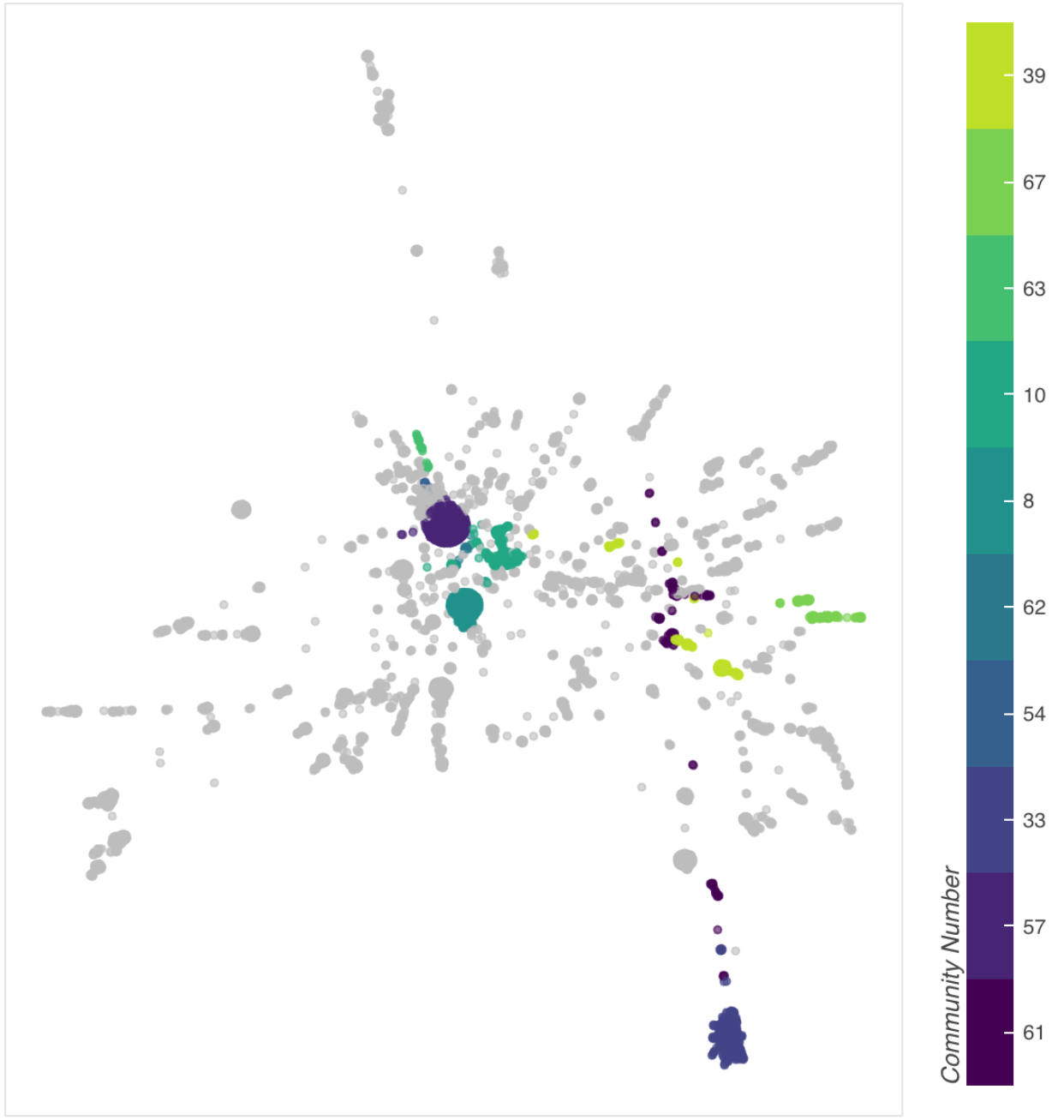}
            \caption{Social Network Analysis}
            \label{fig:sna}
        \end{subfigure}
        \label{fig:sidebyside}
        
    \end{figure}

\end{minipage}

\end{document}